# Anti-vortex dynamics in magnetic nanostripes

Andrew Kunz, Eric C. Breitbach, and Andrew J. Smith

Physics Department, Marquette University, Milwaukee, WI 53233

**Abstract**

In a thin magnetic nanostripe, an anti-vortex nucleates inside a moving domain wall when driven by an in-plane magnetic field greater than the so-called Walker field. The nucleated anti-vortex must cross the width of the nanostripe before the domain wall can propagate again, leading to low average domain wall speeds. A large out-of-plane magnetic field, applied perpendicularly to the plane of the nanostripe, inhibits the nucleation of the anti-vortex leading to fast domain wall speeds for all in-plane driving fields. We present micromagnetic simulation results relating the anti-vortex dynamics to the strength of the out-of-plane field. An asymmetry in the motion is observed which depends on the alignment of the anti-vortex core magnetic moments to the direction of the out-of-plane field. The size of the core is directly related to its crossing speed, both depending on the strength of the perpendicular field and the alignment of the core moments and direction of the out-of-plane field.

**Introduction**

Several proposed devices in magnetic recording, sensing, and logic depend on fast, reliable motion of a domain wall along nanowires.[1,2] In a narrow, thin magnetic nanowire the magnetic moments of the material lie in the plane of the wire and are aligned along the long axis. A transverse domain wall separates the regions of alternating magnetization lying head to head or tail to tail.[3] An external magnetic field applied in the plane of the wire is used to drive the domain wall along the long axis of the wire. It is known that the maximum speed of the domain wall occurs when the applied field is equal to the so-called Walker field, typically on the order of 10 – 20 Oe.[4-6] When the driving field is greater than this critical field, an anti-vortex nucleates inside the domain wall. The anti-vortex stops the forward propagation of the domain wall until the vortex travels across the width of the wire. The nucleation of the anti-vortex decreases the average domain wall speed by an order of magnitude. Recent work has shown that it is possible to suppress the effects of the anti-vortex by applying an additional magnetic field out of the plane of the wire.[7] The out-of-plane field reverses the magnetic moments of the anti-vortex core which changes the direction of the gyrovector and leads to the fast ejection of the nucleated anti-vortex.[8] Reducing the braking effect of the anti-vortex leads to fast domain wall motion for all in-plane driving fields greater than the Walker field. This field can be applied externally or with the addition of a perpendicularly magnetized underlayer.[9]

The out-of-plane (OOP) magnetic field must be strong enough to reverse the magnetic moments in the nucleated anti-vortex.[7] In Fig.1 the average speed of the domain wall is shown as a function of the OOP magnetic field for an in-plane driving field greater than the critical Walker field. Above the critical OOP field there is a sharp increase in the average speed of the domain wall. In the presence of a large OOP field the wall speed approaches the critical speed reached at the Walker field. The inset of Fig. 1 shows the dependence of the critical field on the cross-section dimensions of the nanowire. The out-of-plane critical field increases with wire thickness. An anti-vortex nucleates when the magnetic moments in the domain wall rotate out of the plane of the wire.[7,10] For thin wires the strong shape anisotropy helps to keep the moments in the plane of the wire. However as the thickness of the wire increases, the shape anisotropy decreases and it becomes easier for the magnetic moments to rotate out of the plane of the wire. An increase in the out-of-plane field is necessary to make up for the reduction in the shape anisotropy. The larger OOP field helps to keep magnetic moments aligned in the plane of the wire.

**Simulation**

The Landau-Lifshitz (LL) equation of motion for a magnetic moment *m* is

$$\frac{\partial \vec{m}}{\partial t} = -\gamma \left( \vec{m} \times \vec{H} \right) - \frac{\alpha \gamma}{M_s} \vec{m} \times \left( \vec{m} \times \vec{H} \right) \qquad (1)$$

where $\gamma$ is the gyromagnetic ratio, $\alpha$ is the phenomenological damping parameter, $M_s$ is the saturation magnetization and *H* is the total field experienced by the moment. The first term on the right hand side models the precession of the magnetic moments about the local field, and the second term is responsible for rotating the magnetization into the direction of the field. Micromagnetic simulation of the LL equation of motion is used to model the motion of domain walls in nanowires five microns in length with rectangular cross-sections.[11] The thickness (5 nm – 20 nm) and the width (50 nm – 300 nm) of the wires are varied. The wire is built up from identical cubic grains of uniform magnetization, 2.5 nm or 5.0 nm on edge. The fourth order predictor-corrector time step is less than a picosecond and the damping parameter alpha is 0.008. The materials parameters are those typical of permalloy. The in-plane driving field is applied along +x axis of the wire has a magnitude of 30 Oe in all simulations presented. A 30 Oe field is greater than the Walker field for all wire cross-sections simulated so anti-vortices should always nucleate.

**Results and Analysis**

Above the Walker field the nucleation of an anti-vortex leads to a periodic stepping of the domain wall down the wire. In Fig. 2 the position of the domain wall as a function of time is presented for three different out-of-plane field strengths. The OOP field is always applied along the +z axis. In the absence of the OOP field the wall motion is periodic and symmetric as expected.[10,12] The snapshots in Fig. 2 show the typical domain patterns and the motion of the vortex core. Each of the wall motions plotted begin in state A, and the periodic motion follows the pattern ABCD before repeating. The grey-scale represents the z-component of the magnetic moments and the arrows represent the in-plane magnetization.

The transverse (*y*-direction) core speed is shown in Fig. 3 as a function of the OOP field. A negative field means the core and the OOP field are aligned anti-parallel, and positive means parallel alignment. The transverse speed of the vortex core depends linearly on the strength of the OOP field. We note that while the vortex core crossing time depends linearly on the OOP field strength; the crossing time is independent of the width of the wire.[13] The combination of Figs. 2 and 3 give a complete picture of the core motion. In the absence of an OOP field, and starting with the wall moments in the –*y* direction as shown in state A, the precessional term in the LL equation of motion rotates the wall moments out of the plane of the wire and into the –*z* direction.  Eventually the anti-vortex nucleates at the bottom edge of the wire, with the core moments pointing in the –z direction.  As the core travels upward, it reverses the wall moments behind it, as shown in state B, before exiting at the top edge of the wire.[7,10,14]  The damping term in the LL equation now drives the wall, as shown in state C, along the length of the wire before another anti-vortex nucleates. This time the precessional motion nucleates a core with the magnetic moments aligned in the +z direction and at the top edge of the wire.  The core travels in the downward direction as shown in state D, reversing the magnetization of the domain wall behind it. Once the wall exits at the bottom edge of the wire, the process repeats itself until the wall reaches the end of the wire.  The addition of an out-of-plane field less than the critical value does not stop the periodic motion, but Fig. 2 shows that there are changes in the motion of the core.  When the core moments are aligned in the direction of the OOP field (snapshot D) the core moves more quickly, and when the core moments are anti-parallel (snapshot B) the core moves more slowly.  The width of the plateaus in Fig. 2 show the core crossing time, and the transverse core speed is plotted in Fig 3.  The alignment of the OOP field and the core moments changes the speed of the core.  In addition, the domain wall travels further when the core is going to nucleate anti-parallel to the OOP field (A→B), and it travels a shorter distance when the alignment is parallel (C→D). The total periodicity of the motion remains relatively unchanged from the case in which no OOP field is applied.  The OOP field therefore inhibits but does not fully suppress the nucleation event in the anti-parallel case, but enhances it when parallel.  As shown in Fig. 2, a larger OOP field is necessary to recover fast domain wall motion.

The inset of Fig. 3 shows that the vortex core dimensions are linearly dependent on the alignment of the field and core moments.   (The anti-vortex core is a sharp feature, shown in states B and D in Fig. 2, and we define the diameter of the core to be the distance over which the z-component of the magnetization varies from the bulk value found in the domains and the domain wall.) The Zeeman energy, $-\vec{m}\cdot\vec{H}$, decreases when the core moments and the field are aligned. This reduction in energy leads to larger vortex cores. The domain wall speed is known to depend on the size of the domain wall and using magnetic fields to change the wall size is known to change the wall speed.[4,15] The same mechanism is responsible for the direct relationship between the transverse core speed and diameter.

**Conclusion**

There is an out-of-plane critical field above which domain walls can be driven quickly along wires for in-plane fields greater than the Walker field where vortex nucleation is important. The vortex core is reversed above the OOP critical field. Below this new critical field the OOP field couples with the magnetic moments of the vortex core. This coupling changes the size of the core and causes it to move more quickly across the wire when the alignment is parallel.

**Acknowledgments**

This work was supported by a Research Corporation Cottrell College Science Award and the National Science Foundation, Grant No. DMR 0706194. Thanks to Oleg Tchernyshyov for helpful discussions.

**Figure Captions**

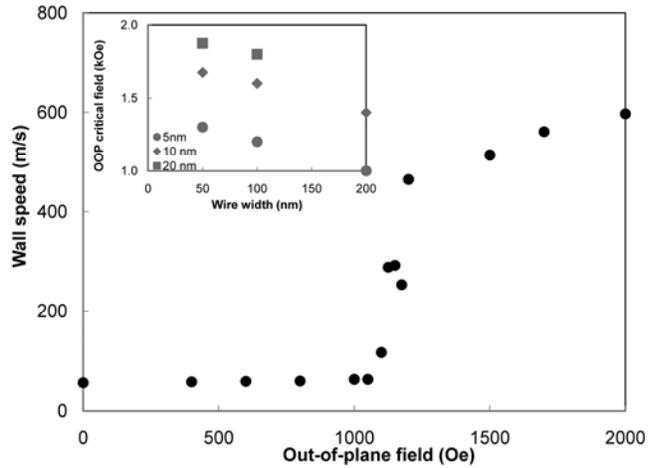

Fig.1. Domain wall speed in a 100 x 5 nm$^2$ cross-section wire as a function of out-of-plane magnetic field when driven by a 30 Oe in-plane field. A sharp increase in wall speed is found at a certain out-of-plane critical field. The inset shows the out-of-plane critical field dependence on the cross-sectional dimensions of the wire.

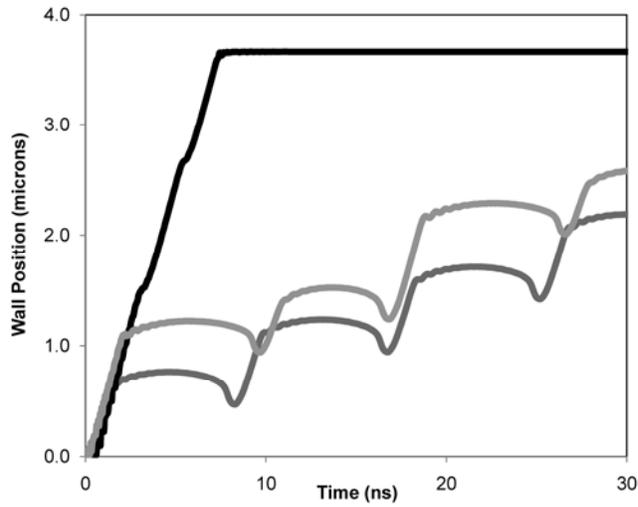

Fig. 2. Plot of the domain wall position as a function of time in a 100 x 5 nm$^2$ cross-section wire for three different out-of-plane field strengths. Below the critical field and asymmetry in the domain wall motion occurs. The magnetic structure of the nanowire is presented in the inset with the grey-scale representing the out-of-plane component of the magnetization. The periodic motion of the domain wall follows the pattern ABCD.

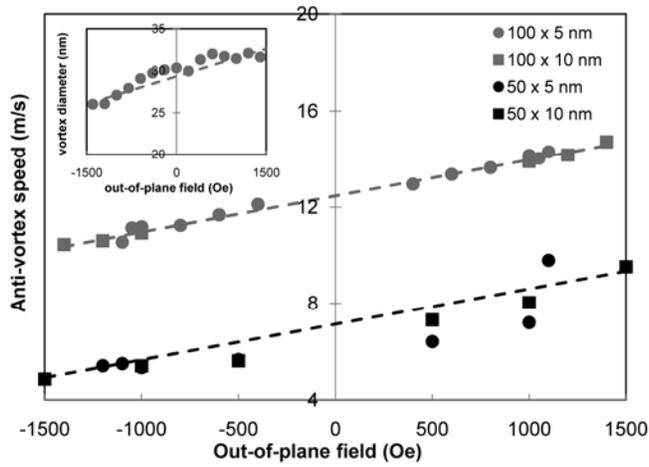

Fig. 3.  Transverse speed of the anti-vortex core as a function of out-of-plane field strength.  Positive fields correspond to parallel alignment of the field and anti-vortex core moments.  The inset shows the corresponding change in anti-vortex core diameter as a function of the out-of-plane field for a 100 x 5 nm$^2$ cross-section wire.